\documentclass[]{elsart}

\usepackage{times}
\usepackage{epsf}
\usepackage[numbers,sort&compress]{natbib}
\usepackage{amsmath}
\usepackage{amsfonts}
\usepackage{amssymb}
\usepackage{bm}
\usepackage{bbm}
\usepackage{color} 
\usepackage{graphicx}
\usepackage{epsfig}
\usepackage{latexsym}
\usepackage{nicefrac}

\def\corresponds{{\lower.2ex\hbox{=}}{\rm\kern-.75em^\triangle}}
\def\succsim{\succ\kern-.9em_\sim\kern.3em}
\def\precsim{\prec\kern-1em_\sim\kern.3em}
\def\slantfrac#1#2{\kern1em^{#1}\kern-.3em/\kern-.1em_{#2}}
\def\lfrac#1#2{{}^{#1\!}\kern-.0em/_{#2}}

\def\buildrel#1\under#2{\mathrel{\mathop{\kern0pt #2}\limits_{#1}}}

\def\e{{\rm e}}

\newcommand{\hf} {\frac{1}{2}}

\newcommand{\uvarphi}{{\underline \varphi}}

\def\mr#1{{\mathrm{#1}}}

\def\cL{{\mathcal L}}

\begin{document}

\begin{frontmatter}

\title{Effective Action and Phase Structure\\
of Multi--Layer Sine--Gordon Type Models}

\author{U. D. Jentschura$^{1}$, I. N\'{a}ndori$^{1,2}$
  and J. Zinn-Justin$^3$}

\address{
$^1$Max--Planck--Institut f\"{u}r Kernphysik, Saupfercheckweg 1,
69117 Heidelberg, Germany \\
$^2$Institute of Nuclear Research of the Hungarian Academy of
Sciences,\\ H-4001 Debrecen, P.O.Box 51, Hungary \\
$^3$ DAPNIA/CEA Saclay,
F-91191 Gif-sur-Yvette, France\\
and Institut de Math\'{e}matiques
de Jussieu--Chevaleret,
Universit\'{e} de Paris VII,
France}

\date{\today}

\begin{abstract}
We analyze the effective action and the
phase structure of $N$-layer sine-Gordon type
models, generalizing the results obtained for the two-layer
sine-Gordon model found in
[I.~N\'{a}ndori, S.~Nagy, K.~Sailer and
U.~D.~Jentschura, Nucl. Phys. B {\bf 725}, 467--492 (2005)]. Besides
the obvious field theoretical interest, the layered sine-Gordon
model has been used to describe the vortex properties of high
transition temperature superconductors, and the extension of the
previous analysis to a general $N$-layer model is necessary
for a description of the critical behaviour of
vortices in realistic multi-layer systems.
The distinction of the Lagrangians in terms of mass eigenvalues
is found to be the decisive parameter
with respect to the phase structure of the $N$-layer models, with
neighbouring layers being coupled by quadratic terms in the field
variables.
By a suitable rotation of the field variables, we identify the periodic modes
(without explicit mass terms) in the $N$-layer structure,
calculate the effective action and
determine their Kosterlitz--Thouless type phase transitions
to occur at a coupling parameter $\beta^2_{{\rm c},N} = 8 N \pi$,
where $N$ is the number of layers (or flavours in terms
of the multi-flavour Schwinger model).
\end{abstract}


\end{frontmatter}

\section{Introduction}
\label{intro} 

The phase structure of generalized sine-Gordon (SG)
type models is known to crucially depend on the periodicity of the interaction 
Lagrangian in the field variable. The ``pure'' SG model is periodic 
in the internal space spanned by the field variable. 
The double-layer sine-Gordon (LSG) 
model~\cite{Fi1979,HeHoIs1995} is characterized by the Lagrangian
\begin{align}
\label{lsg}
{\mathcal L}_{\rm 2LSG} =
{\frac12} \, (\partial \varphi_1)^2 + 
{\frac12} \, (\partial \varphi_2)^2
+ {1\over 2} \, J(\varphi_1 -\varphi_2)^2 
+ U(\varphi_1,\varphi_2)\,,
\end{align}
where $U(\varphi_1,\varphi_2)$ is periodic, but the periodicity 
is broken (partially) by a coupling term between the layers, each 
of which is described by a scalar field. 
Details of the notation are clarified in
Sec.~\ref{def} below. The phase structure of the 
sine-Gordon model is well known~\cite{Ko1974,Amit,sg2,sg3}.
The following
generalization of the SG model,
\begin{align}
\label{clhere}
\cL_{\rm NLSG} = 
{\frac12} \, (\partial {\uvarphi}^\mr{T}) (\partial {\uvarphi})
+ {\frac12} \,J\, \sum_{i=1}^{N-1} (\varphi_i - \varphi_{i+1})^2 +
U(\varphi_1,...,\varphi_N)\,,
\end{align}
belongs to a wider class of massive sine-Gordon type models for $N$ 
coupled Lorentz-scalar fields. For example,
the $N$-layer sine-Gordon model is the bosonized version of the 
$N$-flavour Schwinger model~\cite{Fi1979}. 
The $SU(N)$ Thirring model~\cite{banks,Halpern} is also related to a
suitable generalization of the SG model. 
Periodicity may be broken by explicit mass terms.

\begin{figure}[htb]
\begin{center}
\begin{minipage}{12cm}
\begin{center}
\includegraphics[width=12cm]{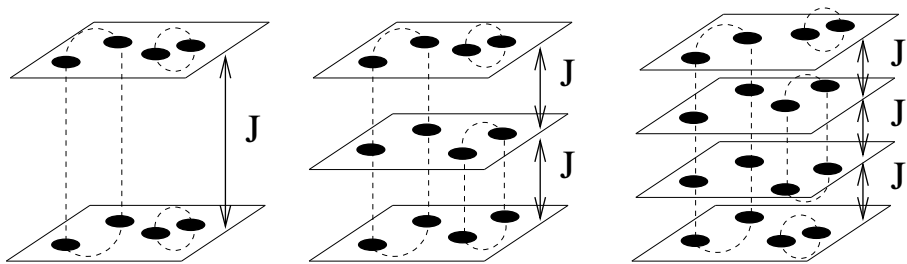}
\caption{
Schematic representation of the multi-layer sine-Gordon model
with $N=2,3,4$ layers which can describe the vortex properties of
layered superconductors. The planes corresponds to 
layered two-dimensional ``sine-Gordon models,'' 
which are coupled by the coupling $J$. The solid discs represent the 
topological excitations of the model, the vortex-antivortex pairs. Two 
such pairs belonging to neighbouring layers can form vortex loops and 
rings due to weak Josephson coupling. The critical behaviour of 
the vortices is found to depend on the number of layers and is again
different in the limit of an infinite number of layers.
\label{layers_vortices}}
\end{center}
\end{minipage}
\end{center}
\end{figure}

The multi-layer sine-Gordon model with $N=2$ layers has been 
proposed as an adequate description of the vortex properties of 
high-$T_{\rm c}$ superconductors which have a strongly anisotropic 
layered structure, and in which the topological excitations in each
two-dimensional superconducting layer are generally thought 
to be equivalent to vortex-antivortex pairs~\cite{pierson_lsg}. Two 
such pairs belonging to neighbouring layers can form vortex loops and 
rings due to the weak Josephson coupling (see Fig.~\ref{layers_vortices}). 
The critical behaviour of the vortices is modified by the sample 
dimensionality; it is different in bulk materials as compared to thin or 
ultra-thin films. 

Recently, models of this type, with only two coupled layers, have 
been analyzed in the framework of the nonperturbative Wegner--Houghton
renormalization group (RG) method which explicitly keeps the periodicity 
in the internal space of the field variable intact~\cite{NaEtAl2005}. 
Here, we
are concerned with a generalization of the previous investigations, by an 
analytic calculation of models with $N$ layers (analyzing the  dependence of 
critical parameters of the ``thickness'' of layered structures).
We decompose the Lagrangians into 
``periodic'' and ``non-periodic'' fields. In general, 
we here refer to a field variable whose self-interaction is 
characterized by a periodic function without an explicit mass term 
as a ``periodic'' mode. Other fields, with an explicit breaking of 
periodicity in the internal space due to a quadratic mass term, will 
be termed ``non-periodic'' modes. 

In Sec.~\ref{def}, we give a 
short overview of the multi-layer SG-type models. 
Starting from a path-integral inspired analysis of the double-layer case
in Sec.~\ref{N=2}, we easily find the generalization 
to the $N$-layer case in Sec.~\ref{genN}.
Confirmation of a central result, to be derived in Sec.~\ref{genN},
is obtained in Sec.~\ref{N=3}, by considering the case
of $N=3$ layers in an alternative functional RG approach.
A summary follows in Sec.~\ref{sum}.

\section{Definition of the multi-flavour massive SG model}
\label{def} 

The general structure of the bare action of a multi-flavour massive 
SG model is~\cite{NaEtAl2005}
\begin{eqnarray}
\label{clb}
\cL &=& {\frac12} (\partial {\uvarphi}^\mr{T}) (\partial {\uvarphi})
+ {\frac12}{ \uvarphi}^\mr{T} {\underline {\underline M}}^2 { \uvarphi}
+ U(\varphi_1,...,\varphi_N),
\end{eqnarray}
where the flavour $O(N)$-multiplet is expressed as a vector of fields,
$\uvarphi= (\varphi_1, \varphi_2, \dots, \varphi_N)^{\rm T}$.
The theory is constructed in $d=2$ spatial dimensions for each 
layer, with a Euclidean metric. 
The global $Z(2)$ discrete symmetry $\uvarphi \to -\uvarphi$
is assumed to leave the Lagrangian invariant. 
The notation $(\partial {\uvarphi}^\mr{T}) (\partial {\uvarphi})$
implies the summation $\sum_{i=1}^d
(\partial_i {\uvarphi}^\mr{T}) (\partial_i {\uvarphi})$
over the dimensions of each layer (in the current paper,
we always have $d=2$).

The interaction term $U(\varphi_1,...,\varphi_N)$ is supposed to be 
periodic in the internal space spanned by the field variables,
\begin{equation}
\label{property2}
U(\varphi_1,...,\varphi_N)= 
U\left(\varphi_1+ \frac{2\pi}{\beta_1},\dots,
\varphi_N+\frac{2\pi}{\beta_N}\right)\,,
\end{equation}
with constant period lengths $\beta_i$.
The mass term in the Lagrangian reads
$\hf \uvarphi^\mr{T} {\underline{\underline M}}^2\uvarphi$,
where the mass matrix $M^2_{ij}$ $(i,j=1,2, ... ,N)$ is symmetric 
and positive semidefinite.
In this paper, we assume the mass matrix to have an 
``interlayer'' structure so that the Lagrangian takes the 
form of Eq.~(\ref{clhere}),
\begin{eqnarray}
\cL_{\rm NLSG} &=& 
{\frac12} \, (\partial {\uvarphi}^\mr{T}) (\partial {\uvarphi})
+ {\frac12} \,J\, \sum_{i=1}^{N-1} (\varphi_i - \varphi_{i+1})^2 +
U(\varphi_1,...,\varphi_N)\,,
\nonumber
\end{eqnarray}
with (initially) $\beta_i = \beta$ (for $i=1,2,...,N$).
An orthogonal transformation of the flavour-multiplet,
$\uvarphi \to{\underline {\underline {\mathcal O}}}\, \uvarphi$,
transforms the model into a similar one with transformed period
lengths in the internal space (which need not all be equal to 
each other). The global $O(N)$ rotation
which connects these bare theories, does not mix the field fluctuations
with different momenta. So, the same global rotation connects the
blocked theories at any given scale, and the scaling laws and
the phase structure therefore are equivalent for 
all models resulting from the orthogonal rotation.

\section{Effective action for the flavour-doublet layered sine-Gordon model}
\label{N=2}

The specialization of Eq.~(\ref{clhere}) to the case of 
$N=2$ layers yields the double-layer sine-Gordon model (LSG), 
whose Lagrangian has been given in Eq.~(\ref{lsg}),
\begin{align}
\label{lsg_periodic}
& {\mathcal L}_{\rm 2LSG} =
{\frac12} \, (\partial \varphi_1)^2 + 
{\frac12} \, (\partial \varphi_2)^2
+ {\frac12} \, J(\varphi_1 -\varphi_2)^2 
\nonumber\\ 
&+ \sum_{n,m=0}^{\infty} 
\left[u_{nm} \cos(n\beta \, \varphi_1)\cos(m\beta \, \varphi_2) +
v_{nm} \sin(n\beta \, \varphi_1)\sin(m\beta \, \varphi_2)
\right]\,.
\end{align}
For $U(\varphi_1, \varphi_2)$, we 
invoke the completeness of the Fourier decomposition of the periodic part.
All running couplings $u_{nm} \equiv u_{nm}(k)$ and 
$v_{nm} \equiv v_{nm}(k)$ are dimensionful (the dimensionless
case will be discussed below). 
The mass matrix reads
\begin{equation}
\label{property3}
{\underline{\underline M}}^2= 
\left( \begin{array}{cc} J &\hspace*{0.5cm} -J\\
-J &\hspace*{0.5cm} J \end{array} \right) \,, \qquad
\det {\underline{\underline M}}^2 \geq 0\,,
\end{equation}
and the mass eigenvalues are $M_+^2=2J>0$ and $M_-^2=0$. 
We now apply a rotation of the field variables
\begin{equation}
\label{rotation2}
\varphi_1 \to \frac{\alpha_1 + \alpha_2}{\sqrt{2}}\,,
\qquad
\varphi_2 \to \frac{\alpha_1 - \alpha_2}{\sqrt{2}}\,.
\end{equation}
The periodic part of the blocked potential at the scale $k$,
\begin{align}
U_k(\varphi_1, \varphi_2) =& \sum_{n,m=0}^{\infty} 
\left[u_{nm}(k) \cos(n\beta \, \varphi_1)\cos(m\beta \, \varphi_2) 
\right. \nonumber\\[2ex]
& \left. +
v_{nm}(k) \sin(n\beta \, \varphi_1)\sin(m\beta \, \varphi_2)\right]
\end{align}
becomes
\begin{align}
& U_k(\alpha_1, \alpha_2) =
\sum_{n,m=0}^{\infty} 
\frac{u_{nm} + v_{nm}}{2} \,
\cos\left[(n-m) \frac{\beta}{\sqrt{2}} \, \alpha_1\right]\,
\cos\left[(n+m) \frac{\beta}{\sqrt{2}} \, \alpha_2\right]\,
\nonumber\\
& \quad  + \sum_{n,m=0}^{\infty} \frac{u_{nm} - v_{nm}}{2} \,
\cos\left[(n+m) \frac{\beta}{\sqrt{2}} \, \alpha_1\right]\,
\cos\left[(n-m) \frac{\beta}{\sqrt{2}} \, \alpha_2\right]\,
\nonumber\\
& \quad  - \sum_{n,m=0}^{\infty} \frac{u_{nm} + v_{nm}}{2} \,
\sin\left[(n-m) \frac{\beta}{\sqrt{2}} \, \alpha_1\right]\,
\sin\left[(n+m) \frac{\beta}{\sqrt{2}} \, \alpha_2\right]\,
\nonumber\\
& \quad  + \sum_{n,m=0}^{\infty} \frac{v_{nm} - u_{nm}}{2} \,
\sin\left[(n+m) \frac{\beta}{\sqrt{2}} \, \alpha_1\right]\,
\sin\left[(n-m) \frac{\beta}{\sqrt{2}} \, \alpha_2\right]
\end{align}
under the rotation. 
It has the general form 
\begin{align}
\label{rotatedLN=2}
& U_k(\alpha_1, \alpha_2) =
\sum_{n,m=0}^{\infty} 
\left[f_{nm} \cos(n b\, \alpha_1)\cos(m b\, \alpha_2) +
h_{nm} \sin(n b\, \alpha_1)\sin(m b\, \alpha_2)\right]\,,
\end{align}
where we identify $b = \beta/\sqrt{2}$ (the notation for $b$
will be frequently used in the following).
We briefly mention the following relations among the 
running couplings, $f_{02} = \hf (u_{11} + v_{11})$, 
$f_{20} = \hf (u_{11} - v_{11})$ and $f_{11} = u_{01} + u_{10}$.
The rotated Lagrangian is 
\begin{align}
& {\mathcal L}_{\rm 2LSG} =
{\frac12} \, (\partial \alpha_1)^2 + 
{\frac12} \, (\partial \alpha_2)^2
+ \frac12 \, M_2^2 \alpha_2^2 
+ U_k(\alpha_1, \alpha_2) \,,
\end{align}
where the mass eigenvalue reads $M_2^2 = J/2$.
We have now disentangled the model into a 
``periodic'' mode $\alpha_1$, for which the Lagrangian 
retains full periodicity in the field variable,
and a non-periodic field $\alpha_2$.

Normally (see, e.g., Ref.~\cite{NaEtAl2005}), 
one assumes the following form for the bare 
Lagrangian  of the double-layer sine-Gordon model,
\begin{align}
\label{bareL}
& {\mathcal L}_{\rm 2LSG} =
\sum_{i=1,2} {\frac12} \, (\partial \varphi_i)^2 
+ {\frac12} \, J(\varphi_1 -\varphi_2)^2 
+ u\,\left[ \cos(\beta \, \varphi_1) + \cos(\beta \, \varphi_2) \right]\,.
\end{align}
We here encounter only the fundamental coupling 
parameter $u = u_{10} = u_{01}$.
Under the rotation (\ref{rotation2}), this Lagrangian 
becomes 
\begin{align}
\label{dothis}
& {\mathcal L}_{\rm 2LSG} =
\sum_{i=1,2} {1\over 2} \, (\partial \alpha_i)^2 
+ \frac12 \, M_2^2 \alpha_2^2 
+ f_{11} \cos(b \alpha_1)\cos(b \alpha_2) \,,
\end{align}
where $f_{11} = 2 u$. It might be worth recalling that
in order to ensure the property of the $\alpha_2 = 0$ field
configuration being a minimum
of the action, we actually have to impose $f_{11} < 0$. 
Based on the argument of the cosine in Eq.~(\ref{dothis}),
one might now be tempted to immediately read off the critical 
value $b^2_{\rm c} = 8 \pi$ for the 
periodic mode $\alpha_1$, which corresponds to 
$\beta_{{\rm c},N=2}^2 = 16 \, \pi$ for the double-layer structure 
as given in Eq.~(\ref{lsg}). This conclusion is especially
tempting because we might have chosen, as the 
bare Lagrangian for the $\alpha_1$-$\alpha_2$-mode configuration, 
a functional form which entails only the couplings $f_{01}$ and $f_{10}$,
\begin{align}
& {\mathcal L}_{\rm 2LSG} =
\sum_{i=1,2} {1\over 2} \, (\partial \alpha_i)^2 
+ \frac12 \, M_2^2 \alpha_2^2 
+ f_{10} \cos(b \alpha_1) + f_{01} \cos(b \alpha_2) \,.
\end{align}
The latter form would have resulted in an immediate decoupling of the 
two fields. However, and somewhat unfortunately,
there is no one-to-one correspondence of the 
rotated couplings $f_{10}$ and $f_{01}$ to the original 
fundamental coupling $u$ which enters into the bare Lagrangian 
as given in Eq.~(\ref{bareL}).

The situation can be remedied, and full confirmation with regard to the 
critical value $\beta_{{\rm c},N=2} = 16 \pi$ can be obtained,
in terms of a
calculational approach inspired by Chap.~9 of Ref.~\cite{ItZu1980},
which leads us to the effective action for the two-layer model.
We restrict 
the discussion to the Fourier mode $f_{11}$ in the rotated  
Lagrangian (\ref{rotatedLN=2}), as in Eq.~(\ref{dothis}),
and calculate the effective Lagrangian for the $\alpha_1$ field. 
We start from the following bare Lagrangian,
\begin{align}
\label{bareLSGrot}
& {\mathcal L}_{\rm 2LSG} =
\sum_{i=1,2} {1\over 2} \, (\partial \alpha_i)^2 
+ \frac12 \, M_2^2 \alpha_2^2 
+ f_{11} \cos(b \alpha_1)\cos(b \alpha_2) \,,
\end{align}
where $f_{11} < 0$. We are interested in the low 
energy behaviour of the model, for $k^2 \ll M_2^2$, 
and therefore use the decomposition 
\begin{align}
\label{dothis2}
& {\mathcal L}_{\rm 2LSG} = {1\over 2} \, (\partial \alpha_1)^2 
+ f_{11} \cos(b \alpha_1)
\nonumber\\
& \quad + {1\over 2} \, (\partial \alpha_2)^2 
+ \frac12 \, M_2^2 \alpha_2^2 
+ f_{11} \cos(b \alpha_1) \left[ \cos(b \alpha_2) - 1 \right]\,,
\end{align}
and expand the cosine into the form
\begin{equation}
\label{expandcos}
\cos(b \alpha_2) \approx 1 - \frac{b^2}{2!} \, (\alpha_2)^2 +
\frac{b^4}{4!} \, (\alpha_2)^4 + {\mathcal O}(\alpha_2)^6\,.
\end{equation}
Here,
the fundamental periodic Lagrangian ${\mathcal L}_2$ reads
\begin{equation}
{\mathcal L}_2(\alpha_1) = {1\over 2} \, 
(\partial \alpha_1)^2 + f_{11} \cos(b \alpha_1)\,.
\end{equation}
In Eq.~(\ref{dothis2}), we have decomposed the Lagrangian
${\mathcal L}_{\rm 2LSG}$ into the fundamental periodic
form for the $\alpha_1$ field, as given by the 
terms ${1\over 2} \, (\partial \alpha_1)^2 + f_{11} \cos(b \alpha_1)$,
a fundamental massive form ${1\over 2} \, (\partial \alpha_2)^2 
+ \frac12 \, M_2^2 \alpha_2^2$ for the $\alpha_2$ field,
and a perturbation given by the last term on the right-hand side
of Eq.~(\ref{dothis2}), which can be integrated out, using the 
Gaussian measure as provided by the fundamental massive form
for the $\alpha_2$ field, to yield the effective action 
for the $\alpha_1$ field.

The effective action $S[\alpha_1]$ is thus given by the
following path integral,
\begin{align}
&\exp(- S[\alpha_1]) = \int {\mathcal D}[\alpha_2] 
\exp \left[- \int\limits_x  {\mathcal L}_2(\alpha_1)\right]\,
\exp\left[- \int\limits_x \left({1\over 2} (\partial \alpha_2)^2 
+ \frac12 M_2^2 \alpha_2^2 \right) \right.
\nonumber\\[2ex]
& \left. - \int\limits_x 
\left( f_{11} \left(- \frac{b^2}{2} \, (\alpha_2)^2 
+ \frac{b^4}{4!} \,  (\alpha_2)^4 + \dots \right) \cos(b \, \alpha_1)
\right)\right]\,,
\end{align}
where $\int_x \equiv \int d^2 x$. 
After the Taylor expansion of the exponential, the effective action
can be written as
\begin{align}
\exp(- S[\alpha_1]) &= 
\exp \left[- \int\limits_x  {\mathcal L}_2(\alpha_1)\right]
\nonumber\\
\int {\mathcal D}[\alpha_2] 
&\exp \left[- \int\limits_x \left({1\over 2} (\partial \alpha_2)^2 
+ \frac12 M_2^2 \alpha_2^2 \right)\right] 
\left[1 + T_1 + T_2 + T_3 \right]\,,
\end{align}
where 
\begin{align}
&T_1 =  f_{11} \int\limits_x \left(
\frac{b^2}{2} \, \cos(b \, \alpha_1) (\alpha_2)^2 
\right)\,, \quad
T_2 = - f_{11} \int\limits_x \left(
\frac{b^4}{4!} \, \cos(b \, \alpha_1) (\alpha_2)^4 
\right)\,,
\nonumber\\
&T_3 = f_{11}^2\, \frac{1}{2!} \,
\int\limits_x \left(
\frac{b^2}{2} \, \cos(b \, \alpha_1) (\alpha_2)^2 
\right)
\int\limits_y \left(
\frac{b^2}{2} \, \cos(b \, \alpha_1) (\alpha_2)^2 
\right)\, .
\end{align}
We now evaluate the terms $T_i$ ($i=1,2,3$).
The  common normalization factor can be expressed as
$Z(0)$, where
\begin{equation}
Z(j) = \int {\mathcal D}[\alpha_2] 
\exp \left[- \int\limits_x \left({1\over 2} (\partial \alpha_2)^2 
+ \frac12 M_2^2 \alpha_2^2 + j \, \alpha_2 \right)\right]\,.
\end{equation}
The result for $\exp(- S[\alpha_1])$ 
as defined in Eq.~(\ref{effLag}) reads
\begin{align}
\exp(- S[\alpha_1]) &= 
Z(0)\,\exp \left[- \int\limits_x  {\mathcal L}_2(\alpha_1)\right]
\nonumber\\
\times & \left[1 + 
\frac{b^2}{2} f_{11} \, \Delta(0) 
\int\limits_x  \cos(b \, \alpha_1) -
\left( \frac{b^2}{2} \right)^2 f_{11} \frac{\Delta(0)^2}{2!} 
\int\limits_x  \cos(b \, \alpha_1) \right. 
\nonumber\\[2ex]
&+ \left( \frac{b^2}{2} \right)^2  f_{11}^2  \frac{\Delta(0)^2}{2} 
\left(\int\limits_x  \cos(b \, \alpha_1)\right)^2 
+ \left( \frac{b^2}{2} \right)^2  \frac{f_{11}^2}{4 M_2^2 \pi} 
\int\limits_x  (\cos(b \, \alpha_1))^2 
\nonumber\\[2ex]
&\left. - \left( \frac{b^2}{2} \right)^2 \frac{f_{11}^2 }{24 M_2^4 \pi}   
\int\limits_x  \left(\partial \cos(b \, \alpha_1)\right)^2
\right]\, .
\end{align}
Here, $\Delta(0)$ involves an ultraviolet divergent tadpole 
integral, which stems from the two-dimensional scalar
propagator
\begin{equation}
\Delta(x - y) = \int \frac{d^2 k}{(2 \pi)^2} \,
\frac{\e^{{\rm i} \bm{k}\cdot (\bm{x} - \bm{y})}}{\bm{k}^2 + m^2}\,.
\end{equation}
A suitable UV regularization for the logarithmically 
divergent quantity $\Delta(0)$ may be introduced
according to Eq.~(2.2) of Ref.~\cite{Amit}. 
Of course, the corresponding tadpole diagram can be 
removed by normal ordering the interaction Lagrangian.
 
The first three terms in this result may be shown 
to exponentiate into a form
\begin{equation}
\label{Leffective}
\cL_{\rm eff}(\alpha_1) = \frac12\, (\partial \alpha_1)^2 + 
f_{11} \, \exp\left(- \frac{b^2}{2}\,\Delta(0)\right) \,
\cos\left(\frac{b}{\sqrt{2}} \,\alpha_1\right)\,,
\end{equation}
leading to a multiplicative renormalization of the coupling 
in comparison to 
${\mathcal L}_2(\alpha_1) = {1\over 2} \,
(\partial \alpha_1)^2 + f_{11} 
\cos\left(\beta \alpha_1/\sqrt{2}\right)$.
This result has the expected and desired structure,
as it should, and confirms the result 
$\beta_{{\rm c},N=2}^2 = 16 \pi$ for $N=2$. 

The complete effective Lagrangian, to order
$f^2_{11}$, reads 
\begin{align}
\label{effLag}
\cL_{\rm eff}(\alpha_1) =& \frac12 (\partial \alpha_1)^2 
\left( 1 + \left( \frac{b^2}{2} \right)^2 \frac{b^2 f^2_{11}}{48 M_2^4 \pi}\,
\sin^2(b \alpha_1) \right)^2 \nonumber\\[2ex]
& + f_{11} \e^{-b^2 \Delta(0)/2} \, \cos(b \alpha_1) +
\left( \frac{b^2}{2} \right)^2 \frac{f^2_{11}}{4 M_2^2 \pi} \,
\cos^2 (b \alpha_1)\,.
\end{align}
The ``one-loop corrections'' of relative order $f^2_{11}$
lead to the generation of higher harmonics ($\cos^2$), which 
are naturally encountered in the full RG flow,
where they leave the phase structure invariant (see, e.g., 
Ref.~\cite{NaEtAl2005}), even if the 
bare Lagrangian as given in Eq.~(\ref{bareL}) contains only 
a single Fourier mode.
The multiplicative modification of the kinetic term
may be reabsorbed into a suitable redefinition of the 
field. We may thus conclude that the sine-Gordon structure (\ref{Leffective}) 
remains valid for the phase structure 
analysis of the periodic field component of the $N$-layer 
sine-Gordon model, even if quantum corrections due to the 
multiplicative interaction of the cosines as given 
in Eq.~(\ref{dothis}) are taken into account.

The universal IR scaling for the non-periodic fields and its connection
to the path integral can now be shown as follows.
We use a slightly more general form for the
bare Lagrangian as compared to (\ref{bareLSGrot}), with couplings
$f_{01}$ and $f_{11}$,
\begin{align}
& {\mathcal L}_{\rm 2LSG} =
\sum_{i=1}^2 {1\over 2} \, (\partial \alpha_i)^2
+ \frac12 \, M_2^2 \alpha_2^2 
+ f_{11} \cos(b \, \alpha_1) \cos(b \, \alpha_2) + 
f_{01} \cos(b \, \alpha_2)\,.
\end{align}
In order to retain the property of the $\alpha_1 = \alpha_2 = 0$
configuration being a minimum, we impose the conditions
$f_{11} < 0$ and $f_{01} < 0$. We expand the 
cosine to second order only,
\begin{equation}
\label{expandcosleading}
\cos(b \alpha_2) \approx 1 - \frac{b^2}{2} \, (\alpha_2)^2 +
{\mathcal O}(\alpha_2)^4\,.
\end{equation}
Using the expansion (\ref{expandcosleading}), it is now
again possible to integrate out $\alpha_2$, leading to 
an effective Lagrangian for $\alpha_1$:
\begin{equation}
\label{eff3}
{\mathcal L}_{\rm eff}(\alpha_1) = {\frac12} \, (\partial \alpha_1)^2 
+ f_{11} \left(1 - \frac{\Delta_{01} \, b^2}{2} \right) 
\cos(b \, \alpha_1) + \frac12\,
\ln(- \Box^2 + M_2^2 + b^2 \vert f_{01} \vert) \,. 
\end{equation}
This representation illustrates that the coupling $f_{01}$ 
effectively shifts the mass term of the non-periodic field $\alpha_2$,
in the IR region ($k \ll M_2$). We recall that 
$|f_{01}| = -f_{01}$ and $|f_{11}| = -f_{11}$.
The term $1 - \half \Delta_{01} \, b^2$
is now easily identified as the first term in the expansion of 
the exponential $\exp(- \half \Delta_{01} \, b^2)$ in powers of 
its argument, confirming the consistency of Eqs.~(\ref{effLag}) 
and~(\ref{eff3}).

The explicit representation in Eq.~(\ref{eff3})
illustrates that the term $\ln(- \Box^2 + M_2^2 + b^2 \vert f_{01} \vert)$ 
now represents a field-independent constant. 
Consequently, we have a vanishing 
RG evolution for the quantity $M_2^2 + b^2 \vert f_{01}\vert $
in the IR. This implies a trivial scaling,
irrespective of $\beta$, for the corresponding dimensionless 
quantities ${\tilde M}^2_2$ and ${\tilde f}_{01}$,
which are related to the corresponding dimensionful quantities 
by the relations~(see Ref.~\cite{NaEtAl2005}) $M_2^2 = k^2 \, {\tilde M}^2_2$
and $f_{01} = k^2 {\tilde f}_{01}$.
The treatment here is easily generalized to 
the three-layer and the $N$-layer structure, and 
confirms that the non-periodic fields have a universal 
scaling in the IR region and do not undergo 
any phase transition (we emphasize that the universal scaling 
is independent of the value of $\beta$).

\section{Generalization of the effective action to $N$ layers}
\label{genN}

The analysis of the $N$-layer structure involves the 
Lagrangian (\ref{clhere}),
\begin{eqnarray}
\cL &=& {1\over 2} \, (\partial {\uvarphi}^\mr{T}) (\partial {\uvarphi})
+ {1\over 2} \, J\,\sum_{i=1}^{N-1} (\varphi_i - \varphi_{i+1})^2 +
U(\varphi_1,...,\varphi_N)\,,
\end{eqnarray}
with initial period lengths $\beta_i = \beta$.
The zero mode of the mass matrix 
\begin{equation}
\frac12\, {\uvarphi}^\mr{T} {\underline {\underline M}}^2 \uvarphi
= {1\over 2} \,J\, \sum_{i=1}^{N-1} (\varphi_i - \varphi_{i+1})^2
\end{equation} 
corresponds to the center-of-mass coordinate of the 
fields,
\begin{equation}
\label{Xcenter}
\alpha_1 = \frac{\varphi_1 + \dots + \varphi_N}{\sqrt{N}}\,.
\end{equation}
The other modes $\alpha_2,\dots,\alpha_N$ become non-periodic.
The transformed Lagrangian reads
\begin{eqnarray}
\cL &=& {1\over 2} \, (\partial {\underline{\alpha}}^\mr{T}) 
(\partial {\underline{\alpha}})
+ {1\over 2} \, \sum_{i=2}^{N} M_i^2 (\alpha_i)^2 +
U(\alpha_1,...,\alpha_N)\,,
\end{eqnarray}
where the transformed periodic potential 
$U(\alpha_1,...,\alpha_N)$ is invariant under the 
transformation $\alpha_1 \to \alpha_1 + 2\pi/b$ 
with $b = \beta/\sqrt{N}$. Because the fields
$\alpha_2, \dots, \alpha_N$ are non-periodic, 
we can assume these to oscillate about their classical
minimum at $\alpha_{i\geq 2} = 0$, following the derivation
leading to Eq.~(\ref{Leffective}). In this first approximation, 
the effective Lagrangian for the periodic field $\alpha_1$
constitutes a generalization of Eq.~(\ref{Leffective}) and reads 
\begin{equation}
\label{LagrangianN}
{\mathcal L}_N(\alpha_1) = \frac12\,(\partial \alpha_1)^2 + 
N u \cos\left(\frac{\beta}{\sqrt{N}} \, \alpha_1\right)\,,
\end{equation}
from which the result 
\begin{equation}
\label{central}
\beta^2_{{\rm c},N} = 8 N \pi
\end{equation}
can be inferred immediately. 
The $N$-layer effective Lagrangian has to be contrasted with
the fundamental SG Lagrangian
\begin{equation}
\label{LagrangianSG}
{\mathcal L}_{\rm SG} = \frac12 \left(\partial \phi\right)^2 + 
u \, \cos(\beta \phi)\,,
\end{equation}
which is {\em a priori} valid in the limit of small $u$.
The increase of $u$, proportional to the number of layers,
therefore means that oscillations of the $\alpha_1$ field about 
the minima of the cosine are severely damped. In general, the strong 
coupling phase of the SG model is characterized by 
a large $\beta$, which translates into fast oscillations of the 
potential as the field variable is changed, and a high tunneling 
probability. A large effective coupling $u \to N u$ suppresses the
tunneling probability and therefore impedes the transition 
to the strong coupling phase. The increase of the critical
parameter $\beta^2_{{\rm c},N} = 8 N \pi$ which separates the two phases 
of the model, is consistent with this trend as the number of layers 
$N$ is increased.

We conclude that as $N \to \infty$, the quantum phase transition
of the layered structure is shifted toward large $\beta$, and 
severely impeded by the increase in the coupling parameter.
For $N \to \infty$, the critical value becomes infinitely large
($\beta_{{\rm c},N} \to \infty$), and the model has only one phase. 
In this continuum limit, the multi-layer model 
can be considered as the discretized version of the three-dimensional
sine-Gordon model (3D-SG) which has recently been shown to have
only one phase, at least 
within in the local potential approximation \cite{sg3}.
The latter observation is entirely consistent with the 
infinite value of $\beta_{{\rm c},N}$ in the continuum limit.

\section{Alternative functional RG approach to $N=3$ layers}
\label{N=3}

In order to obtain additional confirmation with regard
to the validity of the general result (\ref{central}), we 
would like to follow a different route in the current Section,
by analyzing the case of $N=3$ layers in the framework 
of the functional Wegner--Houghton (WH) 
renormalization-group method, whose application to the 
case of $N=2$ layers has already been described in Ref.~\cite{NaEtAl2005}.
Some aspects of the RG study of the 3-layer model have also been
discussed in Ref.~\cite{jpa}.
The specialization of Eq.~(\ref{clhere}) to the case of three layers 
yields the 3-layer sine-Gordon model (3LSG), for which the 
Lagrangian can be written down as follows,
\begin{align}\label{3lsg}
{\mathcal L}_{\rm 3LSG}  =&
{1\over 2} \, (\partial \varphi_1)^2 + 
{1\over 2} \, (\partial \varphi_2)^2 +
{1\over 2} \, (\partial \varphi_3)^2 
+ {1\over 2} \, J(\varphi_1 -\varphi_2)^2 
+ {1\over 2} \, J(\varphi_2 -\varphi_3)^2  
\nonumber\\ 
& + \sum_{n,m,l=-\infty}^{\infty} 
w_{nml} \,
\exp\left({\rm i}\,n \beta\, \varphi_1\right)\,
\exp\left({\rm i}\,m \beta\, \varphi_2\right)\,
\exp\left({\rm i}\,l \beta\, \varphi_3\right)\,.
\end{align}
For the periodic part, we again invoke the completeness of the 
Fourier decomposition. The couplings $w_{nml}$ are dimensionful 
quantities (the transition to the dimensionless case will be 
discussed below). 
We apply the following rotation of the field variables,
$\varphi_1 \to \frac{\alpha_1}{\sqrt{3}} - \frac{\alpha_2}{\sqrt{2}}  
+ \frac{\alpha_3}{\sqrt{6}}$,
$\varphi_2 \to \frac{\alpha_1}{\sqrt{3}}   
- \frac{\sqrt{2}\alpha_3}{\sqrt{3}}$,
$\varphi_3 \to \frac{\alpha_1}{\sqrt{3}} + \frac{\alpha_2}{\sqrt{2}}  
+ \frac{\alpha_3}{\sqrt{6}}$.
The field $\alpha_1$ takes the role of a center-of-mass 
coordinate in the sense of Eq.~(\ref{Xcenter}).
We illustrate the action of this transformation
onto the periodic part of the  potential,
by taking into account the fundamental mode  
of the periodic bare potential, which has a flavour symmetry 
($\varphi_1 \longleftrightarrow \varphi_3$) and reads
\begin{equation}
\label{fundansatz}
U(\varphi_1, \varphi_2, \varphi_3) =
u  \cos(\beta \, \varphi_1) +
u_2 \cos(\beta \, \varphi_2) +
u  \cos(\beta \, \varphi_3) \,,
\end{equation}
with the identifications $u_{100} = u_{001} \equiv u$ and $u_{010}\equiv u_2$.
The fundamental ansatz (\ref{fundansatz}) of the periodic potential becomes
\begin{align}
& U(\alpha_1, \alpha_2, \alpha_3) =
u_2 \,
\cos\left(\frac{\beta}{\sqrt{3}} \, \alpha_1\right)\,
\cos\left(\frac{2\beta}{\sqrt{6}} \, \alpha_3\right)\,
\nonumber\\
& \quad  + 2 u \,
\cos\left(\frac{\beta}{\sqrt{3}} \, \alpha_1\right)\,
\cos\left(\frac{\beta}{\sqrt{2}} \, \alpha_2\right)\,
\cos\left(\frac{\beta}{\sqrt{6}} \, \alpha_3\right)\,
\nonumber\\
& \quad  + 2 u \,
\sin\left(\frac{\beta}{\sqrt{3}} \, \alpha_1\right)\,
\cos\left(\frac{\beta}{\sqrt{2}} \, \alpha_2\right)\,
\sin\left(\frac{\beta}{\sqrt{6}} \, \alpha_3\right)\,
\nonumber\\
& \quad  + 2 u_2 \,
\cos\left(\frac{\beta}{\sqrt{3}} \, \alpha_1\right)\,
\sin\left(\frac{\beta}{\sqrt{2}} \, \alpha_2\right)\,
\sin\left(\frac{\beta}{\sqrt{6}} \, \alpha_3\right)\,.
\end{align}
The general form of the blocked potential is
\begin{align}
\label{rotatedLN=3}
& U_k(\alpha_1, \alpha_2, \alpha_3) =
\nonumber\\
& \quad
\sum_{n,m,l=-\infty}^{\infty} 
j_{nml} \,
\exp\left(\frac{{\rm i}\,n \beta}{\sqrt{3}}\, \alpha_1\right)\,
\exp\left(\frac{{\rm i}\,m \beta}{\sqrt{2}}\, \alpha_2\right)\,
\exp\left(\frac{{\rm i}\,l \beta}{\sqrt{6}}\, \alpha_3\right)\,,
\end{align}
where the $j_{nml}$ are expansion coefficients. The period length 
for the periodic mode $\alpha_1$ is $b_1 = \beta/\sqrt{3}$. 
For the two non-periodic modes, the transformed period lengths read
$b_2 = \beta/\sqrt{2}$ and $b_3 = \beta/\sqrt{6}$. 
The rotated Lagrangian is
\begin{align}
& {\mathcal L}_{\rm 3LSG} =
\sum_{i=1}^3 {1\over 2} \, (\partial \alpha_i)^2 
+ \frac12 \, M_2^2 \alpha_2^2  
+ \frac12 \, M_3^2 \alpha_3^2 
+ U_k(\alpha_1, \alpha_2, \alpha_3) \,.
\end{align}
The mass eigenvalues read $M_2^2 = J$ and $M_3^2 = 3J$.
We have now disentangled the three-layer model into one periodic 
mode $\alpha_1$ and two non-periodic fields $\alpha_2$,  $\alpha_3$.

The functional WH--RG equation for the 
LSG model with $N=3$ layers is a simple generalization of 
the $N=2$ layer equations previously discussed in Ref.~\cite{NaEtAl2005},
\begin{equation}
\label{WHfuncN=3} 
S_{k-\Delta k}[\alpha_1,\alpha_2,\alpha_3] = 
S_k[\alpha_1,\alpha_2,\alpha_3] + {\hbar\over 2} 
{\rm tr} \ln\left(\det S^{ij}_k[\alpha_1, \alpha_2, \alpha_3]\right)\,.
\end{equation}
Here, $S^{ij}_k[\alpha_1,\alpha_2,\alpha_3]$ ($i,j= 1,2,3$) 
denotes the second functional derivative matrix of the blocked action 
with respect to $\alpha_1$, $\alpha_2$ and $\alpha_3$. Again,
the trace is taken over the momentum shell [$k-\Delta k,k$]. 

We now repeat the same steps as in Ref.~\cite{NaEtAl2005}. First, we 
use the local potential approximation (LPA), with the RG evolution 
of the derivative terms being neglected. We start from the 
following general form for the rotated blocked potential for the 
flavour-triplet LSG model,
\begin{align}
V_k=& \frac12 \, M_2^2 \alpha_2^2 + \frac12 \, M_3^2 \alpha_3^2 
\nonumber\\
& + \sum_{n,m,l=-\infty}^{\infty} 
j_{nml} \,
\exp\left({\rm i}\,n b_1\, \alpha_1\right)\,
\exp\left({\rm i}\,m b_2\, \alpha_2\right)\,
\exp\left({\rm i}\,l b_3\, \alpha_3\right).
\end{align}
The generalized WH--RG equation in $d=2$ dimensions,
for three fields $\alpha_{1,2,3}$, reads
\begin{align}
\label{EQWHN=3}
k\,\partial_k V_k &= - \frac{k^2}{4\pi} \, \ln \left(
\frac{[k^2 + V^{11}_k] [k^2 + V^{22}_k] [k^2 + V^{33}_k]}
{ k^2 (k^2 + M_2^2)  (k^2 + M_3^2)} \right. 
- \frac{[V^{13}_k] [k^2 + V^{22}_k] [V^{31}_k]}
{ k^2 (k^2 + M_2^2)  (k^2 + M_3^2)} 
\nonumber\\
& - \frac{[V^{12}_k] [k^2 + V^{33}_k] [V^{21}_k]}
{ k^2 (k^2 + M_2^2)  (k^2 + M_3^2)} 
- \frac{[V^{23}_k] [k^2 + V^{11}_k] [V^{32}_k]}
{ k^2 (k^2 + M_2^2)  (k^2 + M_3^2)} 
\nonumber\\
& + \left. \frac{[V^{12}_k] [V^{23}_k] [V^{31}_k]}
{ k^2 (k^2 + M_2^2)  (k^2 + M_3^2)} 
+ \frac{[V^{13}_k] [V^{21}_k] [V^{32}_k]}
{ k^2 (k^2 + M_2^2)  (k^2 + M_3^2)} 
\right) \,.
\end{align}
Here, $V^{ij}_k= \partial_{\alpha_i}
\partial_{\alpha_j}V_k (\alpha_1,\alpha_2,\alpha_3) $.

We use the ``mass-corrected'' UV approximation the WH--RG equation 
(\ref{EQWHN=3}), which reduces to a set of uncoupled
differential equations for the coupling 
constants of the model 
\begin{align}
& (2+ k\partial_k) {\tilde j}_{nml}(k) = \frac{1}{4\pi}\,
\left( n^2 b_1^2 + \frac{k^2 m^2 b_2^2}{k^2 + M_2^2} + 
\frac{k^2 l^2 b_3^2}{k^2 + M_3^2} \right)
{\tilde j}_{nml} \,,
\end{align}
with the dimensionless quantities $\tilde j_{nml} = k^{-2} j_{nml}$.
The solutions of the RG equations read
\begin{align}
& {\tilde j}_{nml}(k) = {\tilde j}_{nml} (\Lambda)
\left(\frac{k}{\Lambda}\right)^{-2+\frac{b_1^2 n^2}{4 \pi}}
\left(\frac{k^2 + M_2^2}{\Lambda^2 + M_2^2}\right)^{\frac{m^2 b_2^2}{8\pi}}
\left(\frac{k^2 + M_3^2}{\Lambda^2 + M_3^2}\right)^{\frac{l^2 b_3^2}{8\pi}} \,,
\end{align}
where ${\tilde j}_{nml}(\Lambda)$ represents 
the initial condition at the UV cut-off $k = \Lambda$.
In the IR limit ($k\to 0$), the pure non-periodic
modes are relevant couplings $\tilde j_{0ml} \propto k^{-2}$ 
and this is in agreement with IR approximated results obtained 
in Sec.~\ref{N=2}. The periodic modes $\tilde j_{nml}$ for $n>0$
are relevant or irrelevant couplings depending on the
value of $b_1^2$. For the fundamental coupling with $n=1$,
the transition occurs at $b_1^2 = 8\pi$, 
which confirms the value of $\beta_{{\rm c},N=3}^2 = 24\pi$
in view of the relation $b_1^2 = \beta^2/3$
and thus provides additional evidence for the general
result (\ref{central}).

\section{Summary}
\label{sum} 

We have analyzed the phase structure of a general $N$-layer 
sine-Gordon model, as defined in Sec.~\ref{def}, by calculating 
effective actions for the periodic field variables 
(Secs.~\ref{N=2} and~\ref{genN}),
and by considering, as an alternative, functional RG methods
(Sec.~\ref{N=3}).
The multi-layer sine-Gordon model is the bosonized version 
of the multi-flavour Schwinger model, and the flavour-doublet 
(double-layer) sine-Gordon model has been used to describe 
phenomena such as the vortex properties of high 
transition temperature superconductors~\cite{pierson_lsg,kalman,philmag}.

\begin{figure}[htb]
\begin{center}
\begin{minipage}{8cm}
\begin{center}
\includegraphics[width=8cm]{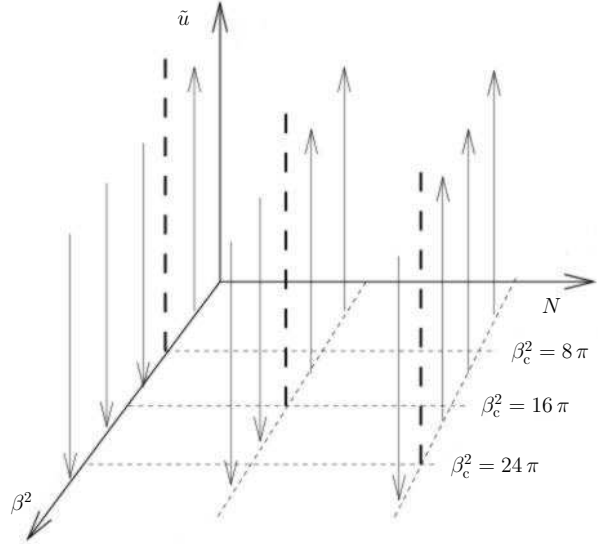}
\caption{Renormalization-group trajectories for the effective coupling 
${\tilde u} = k^{-2}\, u$ as a function of the number of layers $N$,
according to the effective Lagrangian (\ref{LagrangianN}).
The figure illustrates the generalization 
of Fig.~2 of Ref.~\cite{NaEtAl2005} to an arbitrary number of layers.
\label{layers_three}}
\end{center}
\end{minipage}
\end{center}
\end{figure}

In comparison to the previous investigations
(WH approach used in Refs.~\cite{NaEtAl2005,sg3}),
we here take a different route and perform first a 
rotation of the fields,
before calculating the effective action as given in
Eqs.~(\ref{Leffective}) and~(\ref{effLag}).
In the IR, the non-periodic 
mode can be treated perturbatively, by expanding the periodic 
interaction (cosine) in powers of the field. 
The fundamental coupling belonging to any
non-periodic mode is found to have a trivial IR scaling,
and the corresponding dimensionful quantities do not evolve
at all under the RG transformations.
This holds independently of the value of $\beta$, i.e.
independently of the temperature.
In the rotated Lagrangian, only one of
the modes retains a mass term, and the determination
of the general phase structure of the rotated model
therefore becomes possible. 

Generalizing a previous investigation~\cite{NaEtAl2005},
we find that the periodic mode in the $N$-layer structure actually 
undergoes a phase transition at a critical value of 
$\beta^2_{\rm c} = 8 N \pi$ (see Sec.~\ref{genN}).
The effective $N$-layer Lagrangian for the periodic mode 
is given in Eq.~(\ref{LagrangianN}).
In addition to the obvious field-theoretic
interest in related questions, the dependence on the number of 
layers finds a natural application in high-$T_c$ superconductors. 
As has already been stressed, the 
multi-layer sine-Gordon model can be considered as an adequate model
for the vortex behaviour in layered 
superconductors (see Fig.~\ref{layers_vortices}
and Refs.~\cite{pierson_lsg,kalman,philmag}).
The investigations presented here may indicate a possible
explanation for the dependence of the transition
temperature on the thickness of layered systems
(see also Fig.~\ref{layers_three}). 
Experimentally, an increase of the transition temperature of 
high-$T_c$ materials 
with the number of layers has been observed~\cite{experiment}.
Details of the mapping of $\beta^2_{\rm c}$ to the 
transition temperature, and of the transition to a three-dimensional
model for an infinite number of layers, will be presented 
elsewhere. 

Finally, with regard to the connection of the $N \to \infty$ limit
to the three-dimensional case~\cite{sg3},
we reemphasize that the the critical value becomes infinitely large
in this limit ($\beta_{{\rm c},N} \to \infty$ for $N \to \infty$), 
and the model has only one phase.
In this continuum limit, the multi-layer model
can be considered as the discretized version of the three-dimensional
sine-Gordon model (3D-SG), and the general result 
in Eq.~(\ref{central}) is therefore entirely consistent 
with the conclusions of Ref.~\cite{sg3}. In order to 
illustrate the mapping, we observe that the interlayer 
coupling term $J\,(\varphi_i - \varphi_{i+1})^2$,
as given in Eq.~(\ref{clhere}), finds a
natural interpretation as a kinetic term proportional to 
$(\partial \varphi/\partial z)^2$, in the limit $N \to \infty$.
Of course, $z$ here denotes the third spatial direction, 
complementing the $x$ and $y$ integrations relevant for the 
Lagrangian (\ref{clhere}).

%
%
\section*{Acknowledgments}

U.D.J. acknowledges support from 
Deutsche Forschungsgemeinschaft (Heisenberg program), and I.N. 
acknowledges the warm hospitality during a visit to Heidelberg,
and especially the Max--Planck--Institute for 
Nuclear Physics (Heidelberg) for the very fruitful and 
invigorating atmosphere,
as well as numerous discussions at the Institute of Physics, 
University of Heidelberg.

\end{document}